\documentclass{ptapap}
\usepackage{amsmath}
\usepackage{amssymb}

\usepackage{hyperref}

\author{Swayamtrupta Panda}[CFT,LNA,CAMK]
\affil[CFT]{Center for Theoretical Physics, Polish Academy of Sciences, Al. Lotnik\' ow 32/46, 02--668 Warsaw, Poland}
\affil[CAMK]{Nicolaus Copernicus Astronomical Center, Polish Academy of Sciences, Bartycka 18, 00--716 Warsaw, Poland}
\affil[LNA]{Laborat\'orio Nacional de Astrof\'isica - MCTIC, R. dos Estados Unidos, 154 - Na\c{c}\~oes, Itajub\'a - MG, 37504--364, Brazil}

\title{Optical \textmyfont{Fe II} and near-infrared \textmyfont{Ca II} emission in active galaxies}

\newenvironment{myfont}{\fontfamily{lmtt}\selectfont}{\par}
\DeclareTextFontCommand{\textmyfont}{\myfont}

\DeclareRobustCommand{\ion}[2]{%
\relax\ifmmode
\ifx\testbx\f@series
{\mathbf{#1\,\mathsc{#2}}}\else
{\mathrm{#1\,\mathsc{#2}}}\fi
\else\textup{#1\,{\mdseries\textsc{#2}}}%
\fi}

\def\hb{{\sc{H}}$\beta$\/}

\def\feii{Fe {\sc{ii}}}
\def\caii{Ca {\sc{ii}}}
\def\rfe{R$_{\rm{FeII}}$}
\def\rcat{R$_{\rm{CaT}}$}
\def\cat{CaT}

\def\LLEdd{$L\mathrm{_{bol}}$/$L\mathrm{_{Edd}}$}

\def\RL{$R\mathrm{_{H\beta}}-L_{5100}$}

\def\rblr{$R\mathrm{_{BLR}}$}
\def\oi{O {\sc i} $\lambda8446$}
\def\hb{H$\beta$}

\begin{document}

\maketitle

\begin{abstract}
The CaFe Project involves the study of the properties of the low ionization emission lines (LILs) pertaining to the broad-line region (BLR) in active galaxies. These emission lines, especially the singly-ionized iron (\feii{}) in the optical and the corresponding singly-ionized calcium (\caii{}) in the near-infrared (NIR) are found to show a strong correlation in their emission strengths, i.e., with respect to the broad H$\beta$ emission line, the latter also belonging to the same category of LILs. We outline the progress made in the past years that has developed our understanding of the location and the efficient production of these emission lines. We have yet to realize the full potential of \caii{} emission and its connection to the black hole and the BLR parameters which can be useful in - (1) the classification of Type-1 active galactic nuclei (AGNs) in the context of the main sequence of quasars, (2) to realize an updated radius-luminosity relation wherein the inclusion of the strength of this emission line with respect to \hb{} can be an effective tracer of the accretion rate of the AGN, and, (3) the close connection of \caii{} to \feii{} can allow us to use the ratio of the two species to quantify the chemical evolution in these active galaxies across cosmic time. In this paper, we use our current sample and utilize a non-linear dimensionality reduction technique - t-distributed Stochastic Neighbour Embedding (tSNE), to understand the clustering in our dataset based on direct observables. 
\end{abstract}

\section{The CaFe Project - past, present and future}
\begin{figure}
    \centering
    \includegraphics[width=0.5\columnwidth]{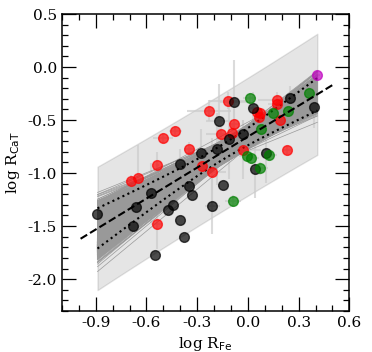}
    \caption{\rfe-\rcat\ relation shown for our sample (in log-scale). Black, red, green and magenta dots correspond to  \citet{persson1988},  \citet{martinez-aldamaetal15, martinez-aldamaetal15b}, \citet{murilo2016} and \citet{2020arXiv200401811M} samples, respectively. Black dotted lines mark the confidence intervals at 95\% for a random 1000 realizations (dark gray lines) of a bootstrap analysis. Light gray shaded region marks the corresponding 95\% prediction intervals bands. Figure courtesy: \citet{martinez-aldama_2021}.}
    \label{fig:correlation}
\end{figure}

The CaFe Project involves the study of the properties of the low ionization emission lines (LILs) pertaining to the broad-line region (BLR) in active galaxies. These emission lines, especially the singly-ionized iron (\feii{}) in the optical (4434-4684 \AA\ blend) and the corresponding singly-ionized calcium (\caii{}) in the near-infrared (NIR, the \caii{} is seen as a triplet $\lambda8498$, $\lambda8542$ and $\lambda8662$ \AA\, and collectively called \cat{}) are found to show a strong correlation in their emission strengths, i.e., with respect to the broad H$\beta$ emission line, the latter also belonging to the same category of LILs (see Figure \ref{fig:correlation}). The origin of this correlation is attributed to the similarity in the physical conditions necessary to emit these lines - especially in terms of the strength of the ionization from the central continuum source and the local number density of available matter in these regions. We refer the readers to \citet{pandaetal2020_paper1,panda2021,martinez-aldama_2021} for a detailed overview on the issue wherein we discuss the robustness of the correlation recovered both from the observational and modelling standpoints. The combined importance of the two species has also been recognized in addition to our findings that the \cat{} being an effective proxy serves to be a better alternative to \feii{}-based \RL{} relation \citep{2021POBeo.100..287M}. This is crucial to address the scatter seen due to the inclusion of newer measurements and sources in the \RL{} relation specifically showing a deviation from the classical two-parameter \RL{} relation \citep{bentz13}. A large subset of these sources are noted to belong to the class of Narrow-line Seyfert Type-1 galaxies (NLS1s) that typically show shorter time delays, and hence a smaller radial distance of the onset of the BLR from the central continuum source (\rblr{}) is deduced for these sources. These smaller \rblr{} values show a marked deviation from the expected \RL{} relation, and addressing this problem is key to our understanding of how these systems evolve and if viable corrections to the classical relation can be made to utilize AGNs as ``standardizable'' cosmological candles \citep{czerny2021_acta}.  In addition to this issue, we find in \citet{martinez-aldama_2021} that the ratio of the \cat{} to \feii{} (justifying the project name - CaFe) is an effective tracer of the chemical evolution of AGNs and can help us probe the co-evolution of the AGN and its host galaxy in more detail. Also, our analysis in \citet{martinez-aldama_2021} has shown the potential of recovering the main sequence for quasars using spectral properties of NIR emission lines, i.e. \oi{} and \cat{}. The main sequence of quasars \citep{bg92,sul00,mar18} is an outcome of a classical linear dimensionality reduction technique - principal component analysis\footnote{wherein the dataset is transformed to be represented in sets of eigenvectors and their corresponding eigenvalues} (PCA), and the primary eigenvector relates to the correlation seen between the full-width at half maximum (FWHM) of broad \hb{} and the optical \feii{} strength (i.e., the \rfe{}\footnote{\rfe{} is the ratio of the optical \feii{} emission within 4434-4684 \AA\ normalized to the \hb{} emission, and, \rcat{} is the ratio of the NIR \cat{} emission normalized also to the same \hb{} emission.}). As the correlation is drawn using broad emission lines seen in a quasar spectrum, this applies to only Type-1 AGNs - those where the distant observer has an almost uninterrupted view of the central engine (i.e. the accreting supermassive black hole) and its immediate surroundings, e.g. BLR that gives rise to these broad emission lines. In \citet{martinez-aldama_2021}, we find that the FWHM(O {\sc i}) recovered for our sample shows an almost 1-to-1 relation with FWHM(\hb{}). This result compounded with the strong \rfe{}-\rcat{} relation suggests the presence of a similar correlation between the FWHM(O {\sc i}) and \rcat{}. This is shown in Figure 1 of \citet{martinez-aldama_2021}. 

\section{How to assess high-dimensional data properly}

Visualization of high-dimensional data is crucial to improve our understanding of how a system behaves, especially factoring in the importance of observables that drive the dataset. Datasets in astronomy are no different and with the huge influx of data that is already upon us, more intelligent classification and clustering schemes are needed, in addition to making them human-readable. We thus rely upon dimensionality reduction techniques - where an n-dimensional dataset can be transformed and represented in a hyper-plane consisting of only a few \textit{hyper}-dimensions. Dimensionality reduction aims to preserve as much of the significant structure of the high-dimensional data as possible in the low-dimensional map.

In the case of principal component analysis \citep[PCA, see][]{Jolliffe2011}, a classical \textit{linear} dimensionality reduction technique\footnote{Over the last few decades, a variety of techniques for the visualization of such high-dimensional data have been proposed, many of which are reviewed by \citet{Oliveira2003} and \citet{baron_2019}.}, these hyper-plane dimensions are constructed using singular value decomposition (SVD), such that the axes generated are orthogonal to each other. The original dataset is then represented in this hyper-plane where they are organised based on their respective contributions to these new dimensions\footnote{In \citet{martinez-aldama_2021}, we explored the observational properties of our sample and performed a principal component analysis, where $81.2\%$ of the variance can be explained by the first three principal components drawn from the full width at half maximum (FWHMs), continuum luminosity of the sources, and the equivalent widths (EWs) of the emission lines considered.  The first principal component (PC1) is primarily driven by the combination of black hole mass and luminosity with a significance over $99.9\%$, which in turn is reflected in the strong correlation of the PC1 with the Eddington ratio. We also discuss the biases introduced in PCA due to the use of input observables that are either derived from existing observables, i.e. black hole mass is derived from two basic observables, FWHM of the broad emission line observed in the AGN spectrum and distance of the BLR from the continuum source.}. But, there is a major drawback with PCA or PCA-based methods, i.e. the new axes are a linear combination of original observables. Thus, no cross-terms are involved when constructing the new dimensions. Therefore users of PCA need to be aware of this assumption and refrain from injecting datasets where prior inter-dependencies are noticed. Yet, PCA remains one of the most sought after methodologies, primarily due to its easy-to-understand algorithm. 

For datasets with inter-dependent observables, one proceeds to look for other methods, for example, t-distributed Stochastic Neighbour Embedding \citep[tSNE,][]{JMLR:v9:vandermaaten08a} which is a non-linear dimensionality reduction technique that aims to preserve the local structure of data. The foundation of tSNE is based on minimizing a cost function that is dependent on the nearness of a data point from a defined cluster center. The algorithm orders the dataset by assigning a rank to each data point and then organizing them on the hyper-plane. In this way, the dataset is represented based on similarity which is estimated based on the input observables for each data point. tSNE employs a heavy-tailed distribution in the low-dimensional space to alleviate both the crowding problem and the optimization problem \citep[see][for a detailed overview of the technique]{JMLR:v9:vandermaaten08a}.

\section{Methods and Analysis}

Our analysis is based on the observational properties of \hb, optical \feii{} and NIR \cat{} triplet collected from \citet{persson1988}, \citet{martinez-aldamaetal15, martinez-aldamaetal15b}, \citet{murilo2016} and \citet{2020arXiv200401811M}.  The full sample includes 58 objects with 42.5 $<$ log L$_{\rm opt}$ (5100 \AA) $<$ 47.7 at $0.01<z<1.68$. Table A1 in \citet{martinez-aldama_2021} reports the properties of the each source in the sample such as redshift, optical (at 5100 \AA) and NIR (at 8542 \AA) continuum luminosity, the flux ratios \rfe\ and \rcat, as well as the equivalent width (EW) and Full-Width at Half Maximum (FWHM) of \feii{}, \hb, \cat\ and O {\sc i} (see Table A1 in \citealt{martinez-aldama_2021}).

\begin{figure}[!h]
    \centering
    \includegraphics[width=0.75\columnwidth]{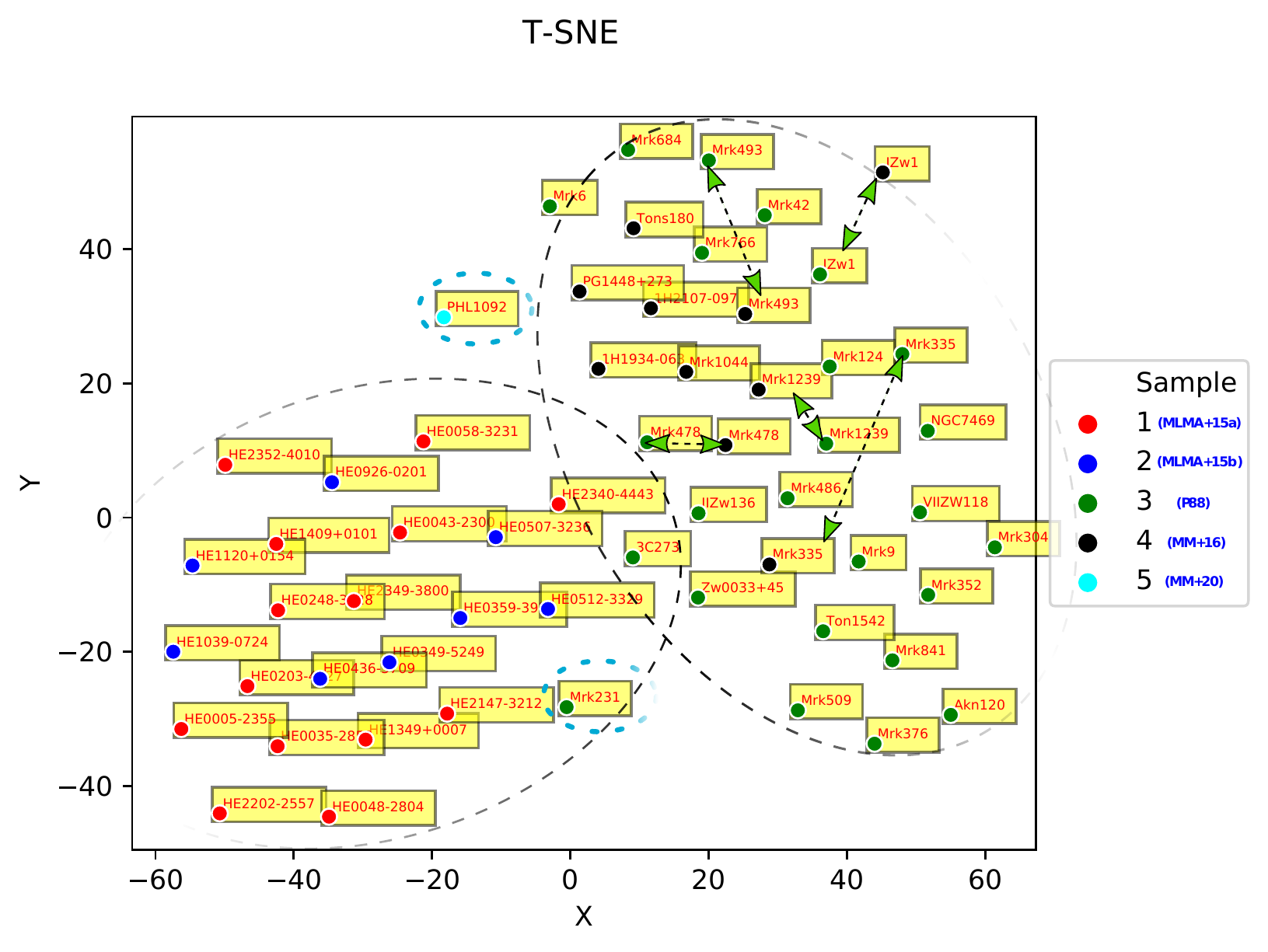}
    \caption{Two-dimensional hyper-plane generated using the tSNE. Green, red, blue, black and cyan dots correspond to  P88 - \citet{persson1988},  MLMA+15a, MLMA+15b - \citet{martinez-aldamaetal15, martinez-aldamaetal15b}, MM+16 - \citet{murilo2016}, and, MM+20 - \citet{2020arXiv200401811M} samples, respectively. The dashed ellipses are shown for illustrative purposes to highlight the dichotomy in the sample and mark sources of interest. The green arrows mark the movement of the five sources which were observed twice at different epochs.}
    \label{fig:tsne_XY}
\end{figure}

In this work, we utilize the \textmyfont{PYTHON} implementation of the tSNE\footnote{\href{https://scikit-learn.org/stable/modules/generated/sklearn.manifold.TSNE.html}{https://scikit-learn.org/stable/modules/generated/sklearn.manifold.TSNE.html}} to analyze the clustering in our sample accounting for the aforementioned spectral properties for each source, specifically, the L$_{\rm opt}$, FWHMs of \feii{}, \hb, \cat\ and O {\sc i}, and the EWs for the same set of emission lines. This prescription of choosing only the direct observables was shown to provide clear evidence of the primary driver of our sample reducing the biases arising from inter-dependencies due to the inclusion of derived parameters, like BH mass or Eddington ratio (\LLEdd{}). The 2D hyper-plane generated using the tSNE is shown in Figure \ref{fig:tsne_XY}. The sources are categorized based on the papers that reported them.

\begin{figure}[!h]
    \centering
    \includegraphics[width=0.75\columnwidth]{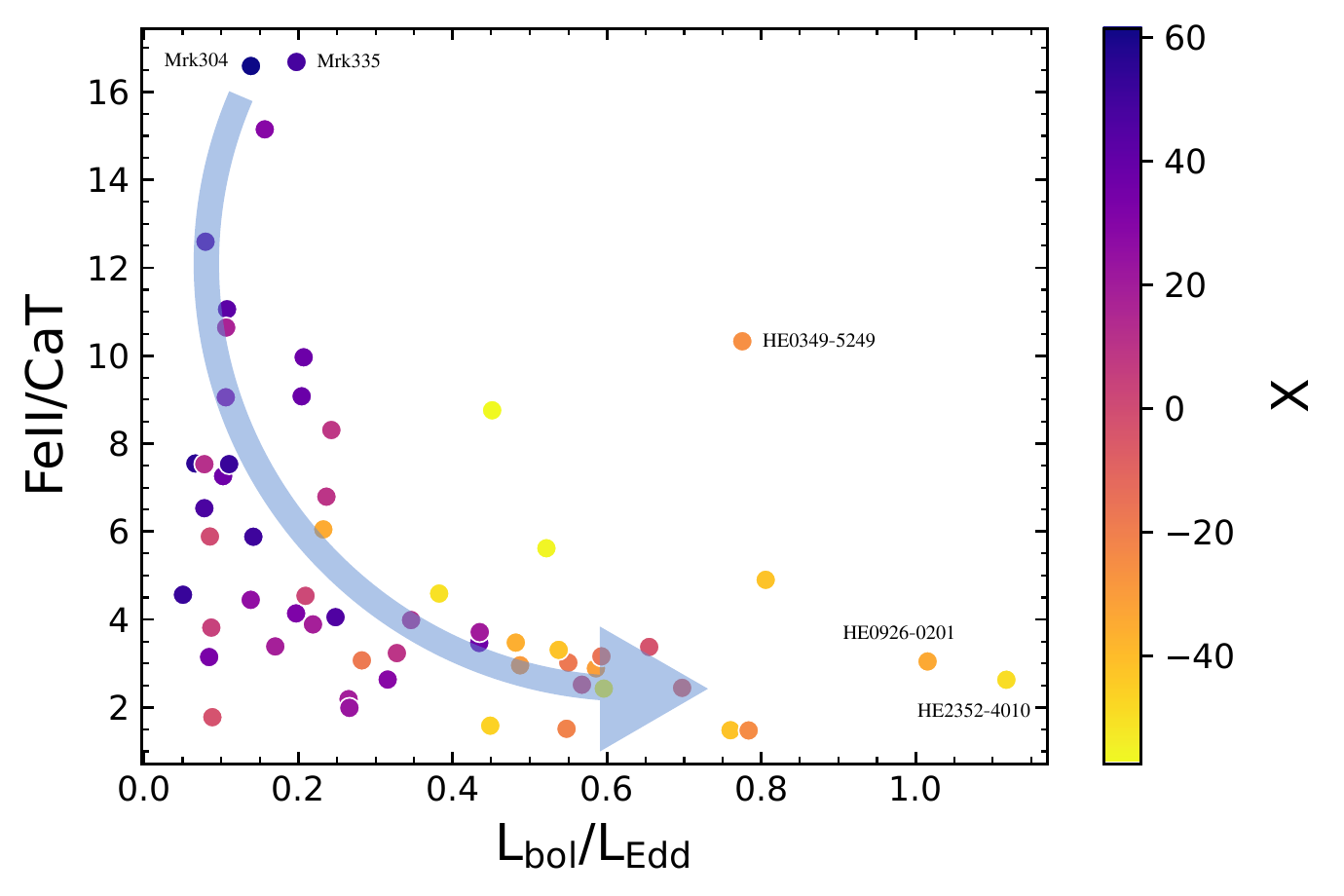}
    \caption{ \feii{}/\cat{} versus Eddington ratio (\LLEdd{}) for the sources in our sample. The color-scale represents the primary tSNE hyper-plane axis, X. The arrow marks the increase in the \LLEdd{}. Sources at extremes are marked.}
    \label{fig:FeCa_Edd}
\end{figure}

\section{Results \& Conclusions}


In this work, we utilize a non-linear dimensionality reduction technique - tSNE, to understand better the organization of our current sample from our CaFe Project. The preliminary testing has revealed a dichotomy in our sample when visualized in a 2D tSNE-based hyper-plane - the overall sample is composed of two clusters driven by their observed optical luminosities and black hole masses (and hence, their Eddington ratios). The analysis has also revealed the movement of sources in this hyper-plane which we observed more than once. As was found earlier from our direct correlation in \citet{pandaetal2020_paper1,martinez-aldama_2021}, the source PHL1092 stands out as an outlier. Here, in addition to that, we notice that this source doesn't belong to either of the clusters. This can be due to the prominence of strong soft X-ray excess seen in its spectrum. This may explain the recovery of high values for \rfe{} (= 2.576$\pm$0.108) and \rcat{} (= 0.839$\pm$0.038) for this source. Another source, Mrk335, is also of potential interest, as it seems to have jumped to the high-luminosity cluster (red and blue dots) albeit belonging to a sub-sample whose remaining sources are located in the low-luminosity cluster (green and black dots).

We explore further the results from the tSNE analysis which reveals a strong and clear connection between the ratio \feii{}/\cat{} (this ratio has been studied in \citet{martinez-aldama_2021} as a tracer of the chemical evolution in galaxies hosting accreting super-massive black holes) and the Eddington ratio (\LLEdd{}) for the sources in our sample - sources with lower \feii{}/\cat{} belong to the high-luminosity class of objects that typical have relatively large \LLEdd{} (see Figure \ref{fig:FeCa_Edd}). This result needs to be tested more carefully.

The advent of newer, deeper surveys including the James Webb Space Telescope (JWST, \citealt{jwst2006}) and Maunakea Spectroscopic Explorer (MSE, \citealt{2019BAAS...51g.126M}), will lead to populating our sample by orders of magnitudes and firmly confirming these aforementioned avenues.

\acknowledgements{The project was partially supported by the Polish Funding Agency National Science Centre, project 2017/26/\-A/ST9/\-00756 (MAESTRO  9), MNiSW grant DIR/WK/2018/12 and acknowledges partial support from CNPq Fellowship (164753/2020-6).}

\bibliographystyle{ptapap}
\bibliography{panda1}

\end{document}